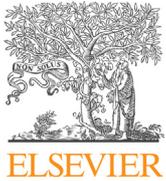
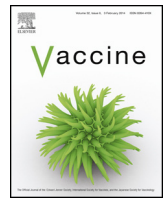
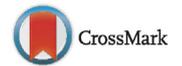

# Vaccination with *Leishmania mexicana* LPG induces PD-1 in CD8+ and PD-L2 in macrophages thereby suppressing the immune response: A model to assess vaccine efficacy

M.B. Martínez Salazar [a], J. Delgado Domínguez [a], J. Silva Estrada [a], C. González Bonilla [b], I. Becker [a],*

[a] *Departamento de Medicina Experimental, Facultad de Medicina, Universidad Nacional Autónoma de México, Hospital General de México, Dr. Balmis 148, Colonia Doctores, 06726, Mexico D.F., Mexico*
[b] *Laboratorios de Vigilancia e Investigación Epidemiológica, Coordinación de Vigilancia Epidemiológica, Instituto Mexicano del Seguro Social, México D.F., Mexico*

**ABSTRACT**

*Leishmania* lipophosphoglycan (LPG) is a molecule that has been used as a vaccine candidate, with contradictory results. Since unsuccessful protection could be related to suppressed T cell responses, we analyzed the expression of inhibitory receptor PD-1 in CD8+ and CD4+ lymphocytes and it is ligand PD-L2 in macrophages of BALB/c mice immunized with various doses of *Leishmania mexicana* LPG and re-stimulated *in vitro* with different concentrations of LPG. Vaccination with LPG enhanced the expression of PD-1 in CD8+ cells. Activation molecules CD137 were reduced in CD8+ cells from vaccinated mice. *In vitro* re-stimulation enhanced PD-L2 expression in macrophages of healthy mice in a dose-dependent fashion. The expression of PD-1, PD-L2 and CD137 is modulated according to the amount of LPG used during immunization and *in vitro* re-stimulation. We analyzed the expression of these molecules in mice infected with $1 \times 10^4$ or $1 \times 10^5$ *L. mexicana* promastigotes and re-stimulated *in vitro* with LPG. Infection with $1 \times 10^5$ parasites increased the PD-1 expression in CD8+ and diminished PD-L2 in macrophages. When these CD8+ cells were re-stimulated *in vitro* with LPG, simulating a second exposure to parasite antigens, PD-1 expression increased significantly more, in a dose dependent fashion. We conclude that CD8+ T lymphocytes and macrophages express inhibition molecules according to the concentrations of *Leishmania* LPG and to the parasite load. Vaccination with increased amounts of LPG or infections with higher parasite numbers induces enhanced expression of PD-1 and functional inactivation of CD8+ cells, which can have critical consequences in leishmaniasis, since these cells are crucial for disease control. These results call for pre-vaccination evaluations of potential immunogens, specifically where CD8 cells are required, since inhibiting molecules can be induced after certain thresholds of antigen concentrations. We propose that the analysis of PD-1 and PD-L2 are useful tools to monitor the optimal dose for vaccination candidates.

## 1. Introduction

*Leishmania* lipophosphoglycan (LPG), one of the principal molecules of the parasite, modulates the immune response. LPG is a ligand for TLR2 in NK cells regulating their IFN-γ and TNF-α production [1]. In mast cells and macrophages LPG modulates TLR2 and protein kinase-alpha (PKC-α), respectively [2,3]. CD4+ lymphocytes define *Leishmania* infections, where a Th-1 aids parasite control and Th-2 response favors disease progression in mouse models [4]. A major role in the defense against *Leishmania* is played by CD8+ cells, both by IFN-γ production and cytotoxicity [5–7]. Activation of CD8+ and CD4+ lymphocytes is regulated by PD-1, an inhibition receptor whose two ligands are PD-L1 (B7-H1) and PD-L2 (B7-DC) [8,9]. The recognition of PD-1 by either ligand leads to a functional exhaustion of CD8+ lymphocytes, characterized by reduced proliferation, the absence of cytokine production and a failure to exert cytotoxicity [10,11]. Yet some evidence also suggests that these molecules modulate CD8+ cells during *Leishmania mexicana* infections. A reduction of CD8+ lymphocytes has been observed in patients with diffuse cutaneous leishmaniasis (DCL), infected with *L. mexicana*. These cells showed enhanced expression of PD-1 and were hampered in their effectors mechanisms, being non-responsive in their cytokine production and showing limited cytotoxicity, when confronted with autologous *Leishmania*-infected macrophages [12,13]. In a model of experimental chronic visceral leishmaniasis caused by *Leishmania donovani*, CD8$^+$ cells were found to show phenotypic markers of functional exhaustion [14]. PD-L2 is a ligand for PD-1 displayed on dendritic cells and macrophages, both of which are host cells for *Leishmania* [9]. For protection against *Leishmania* infections, a fine-tuned regulation leading to CD8$^+$ cell activation is crucial, which includes the induction of co-stimulatory signals and activation molecules such as CD137, favoring cell survival, and the inhibition of PD-1 to avoid cellular anergy.

* Corresponding author. Tel.: +52 55 56232674; fax: +52 55 57610249.
E-mail address: becker@unam.mx (I. Becker).

LPG has been widely used as a vaccine candidate against leishmaniasis, with contradicting results. Thus, subcutaneous immunization with LPG has failed to protect BALB/c mice against *Leishmania amazonensis* infections, exacerbating the disease by enhanced TGF-β and IL-10 production [15]. The administration of anti-LPG antibodies or the intranasal administration of LPG was shown to revert this effect [16].

One of the main pitfalls during vaccination schemes that end unsuccessfully is the use of given antigen concentrations, without previous analysis as to whether this immunogen induces inhibitory or activation molecules. Furthermore, the diverse protection models vary widely in parasite numbers used during the infection challenge, which also accounts for possible contradicting results. To gain insight into the unpredictable outcomes of the different LPG vaccination models, we analyzed if different *L. mexicana* LPG concentrations showed diverse modulation of the inhibitory PD-1 molecule expression in T lymphocytes and PD-L2 expression in macrophages. Additionally we analyzed the influence of the parasite load on the expression of these molecules.

## 2. Material and methods

### 2.1. Animals

Male BALB/c mice aged to 6–8 weeks were bred and housed at the animal facilities of the Departamento de Medicina Experimental of the Medical Faculty, UNAM, following the National Ethical Guidelines for Animal Health NOM-062-ZOO-1999 and the guidelines recommended for animal care by the Ethical Committee of the Medical School of the UNAM.

### 2.2. Leishmania mexicana culture

*L. mexicana* parasites were grown in RPMI-1640 medium (Life Technologies Laboratories, Gaithersburg, MA, USA), supplemented with 10% heat-inactivated FBS at 28 °C. Metacyclic promastigotes were harvested at late log phase (5 day culture).

### 2.3. Lipophosphoglycan purification

Lipophosphoglycan was purified from *L. mexicana* as previously described [1].

### 2.4. Vaccination and infection

For vaccination assays, LPG was suspended in sterile PBS at a final concentration of 1 µg/µL. Mice received three subcutaneous injections (insulin syringe, needle 31 G BD) in the dorsum containing 10 or 100 µg of LPG or 100 µL PBS as control, at a 15 day interval. The protection assay was carried out 20 days after the last vaccination. Mice were infected subcutaneously (insulin syringe, needle 31 G BD) with $1 \times 10^5$ *L. mexicana* promastigotes in the ear dermis. The lesion was measured weekly with a Vernier. For infection analysis, non-vaccinated mice were infected with $1 \times 10^4$ or $1 \times 10^5$ promastigotes and sacrificed prior to ulceration of the lesions.

### 2.5. Peritoneal cells

Mice were sacrificed by cervical dislocation. The peritoneal cavity was infused with 10 mL of cold sterile PBS pH 7.4 and lightly massaged. The peritoneal fluid was collected and centrifuged at $800 \times g$ for 10 min at 4 °C. The cells were cultured 2 h in RPMI 1640 (supplemented with 100 U/mL penicillin and 100 IU/mL streptomycin) containing 10% (v/v) heat-inactivated FBS (RPMI–FBS) at 37 °C with 5% $CO_2$. Macrophages ($1 \times 10^6$/mL) were maintained in 24-well cell culture plates (Corning). Different LPG concentrations (1, 5 or 10 µg) were added, and a negative control contained only culture medium. After 24 h the cells were harvested and analyzed by flow cytometry.

### 2.6. Splenocyte purification

The spleen was aseptically removed and placed in a Petri dish containing cold PBS. The tissue was disrupted in a 100 µm nylon cell strainer (BD Falcon) and the isolated cells were centrifuged at $800 \times g$ for 10 min at 4 °C. Cells were separated by Ficoll–Hypaque gradient (Sigma) and mononuclear cells were washed twice with PBS and placed in 6-well plates (Corning) at $5 \times 10^6$ cells per well and stimulated with 1, 5 or 10 µg *L. mexicana* LPG during 24 h.

### 2.7. Flow cytometry

The extracellular expression of PD-1, CD137, PD-L2 and PD-L1 was analyzed in stimulated or non-stimulated peritoneal macrophages and mononuclear cells ($1 \times 10^6$ cells/mL) were suspended in 100 µL FACS buffer (BD Biosciences cat. 342003) containing CD16/32 antibodies for 10 min on ice. After washing, cells were stained in 50 µL FACS buffer containing fluorochrome-labeled antibodies specific for CD3e (BD Pharmingen cat. 553066), CD8a (BD Pharmingen, cat. 551162), CD4 (BD Pharmingen, cat. 552775), CD137 (BD Pharmingen cat. 558976), F4/80 (Biolegend, cat. 122615), PD-1 (Biolegend, cat. 135205), PD-L1 (Biolegend, cat. 124311), PD-L2 (Biolegend, cat. 107205) or appropriate isotype controls, for 20 min on ice. Cells were then washed twice, fixed in 2% paraformaldehyde and analyzed using a FACSCanto II flow cytometer equipped with DIVA software (BD Biosciences, USA).

### 2.8. Statistical analysis

All data are expressed as mean ± SD (standard deviation of the mean). Comparisons between experimental groups were performed using Mann–Whitney *U*-test. A value of $p < 0.05$ was considered statistically significant, using Prism 5 for Mac OS X®. Three or more independent experiments were analyzed for three mice per group.

## 3. Results

### 3.1. Vaccination with LPG induces exacerbation and progression of L. mexicana infection

Our group previously demonstrated that LPG exerts an immunomodulatory effect on different cells of the immune response [1–3]. We were therefore interested in analyzing whether this molecule could confer protection against *L. mexicana* infections. BALB/c mice were vaccinated with 10 µg *L. mexicana* LPG.

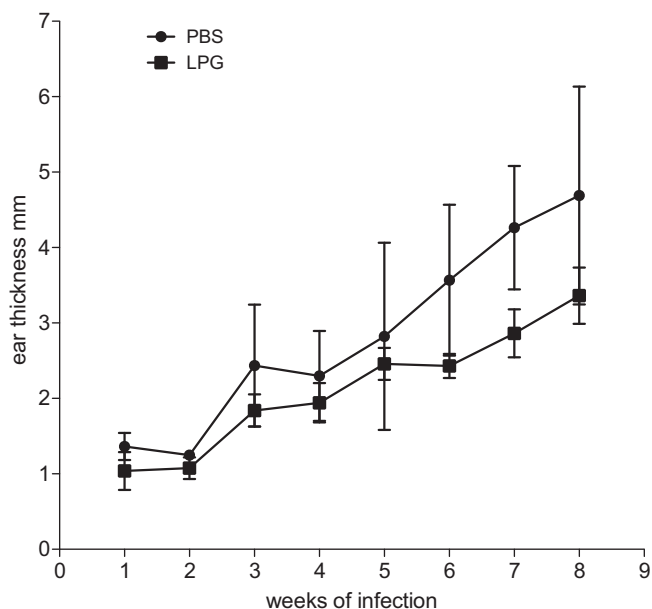

**Fig. 1.** LPG vaccination does not protect mice from *L. mexicana* infection. Male BALB/c were vaccinated at day 0, 15 and 30 with 10 μg LPG (■) or inoculated with PBS (●) and challenged in ear dermis at day 50 with $1 \times 10^5$ *L. mexicana* promastigotes. The infection was followed for 8 weeks and ear lesion was measured weekly. ($n = 4$).

Twenty days after the third immunization, mice were challenged in ear dermis with $1 \times 10^5$ *L. mexicana* promastigotes and the infection was followed throughout 8 weeks. Once the inflammation was detectable, the lesion was measured weekly with a Vernier. Control mice were injected with 10 μL PBS. The ear dermal lesions appeared first in non-vaccinated mice around the third week. Lesions of mice vaccinated with LPG appeared around the fourth week. Throughout the course of the infections, both groups of mice showed similar inflammatory lesions (Fig. 1). After 6 weeks, only the vaccinated mice began to show dissemination of the parasite, forming nodules in the contralateral earlobe, paws and nose, simulating diffuse cutaneous leishmaniasis found in humans (data not shown). Once the disease disseminated in vaccinated mice, the inflammatory lesions in their earlobes tended to evolve slower after 6–7 weeks of infection, as compared to non-vaccinated mice (Fig. 1). It remains to be analyzed whether dissemination increases overall *Leishmania* numbers that possibly induce inhibitory molecules on inflammatory cells, thereby diminishing the inflammation yet not the disease progression. These data show that vaccination with LPG induces a more rapid dissemination of the parasites.

### 3.2. Macrophages infected in vitro with L. mexicana and stimulated with LPG over-express PD-L2 but show no changes in PD-L1 expression

We studied the modulation exerted by *in vitro* stimulation of macrophages from healthy mice with LPG (1, 5 or 10 μg) and analyzed the ligands of regulatory molecules of T cells in macrophages. Stimulation with 1 μg LPG led to an increased PD-L2 expression, yet when the challenge was augmented to 5 μg, the PD-L2 expression significantly increased (3-fold) whereas stimulation with 10 μg only slightly enhanced the expression (2-fold), which was not different from non-stimulated controls (Fig. 2A). These results suggest that LPG is capable of regulating the interaction between T lymphocytes and macrophages by inducing PD-L2 in a dose-dependent fashion.

Furthermore we analyzed whether *in vitro* infection of macrophages could regulate the expression of these inhibitory molecules. Peritoneal macrophages were infected with *L. mexicana* promastigotes in a ratio 1:10 (cells:parasites). In one group, *Leishmania* promastigotes combined with 5 μg LPG were used to infect macrophages. The cells were stained with antibodies against F4/80, PD-L1 and PD-L2. PD-L1 expression decreased slightly in macrophages infected with *Leishmania* promastigotes (Fig. 2B). In contrast, PD-L2 was up-regulated (2.4-fold) in macrophages infected with *Leishmania* combined with LPG, as compared to non-infected cells (Fig. 2B). In conclusion, LPG stimulation seems to have a more potent effect to induce PD-L2 in peritoneal macrophages, as compared to the infection with *L. mexicana* alone.

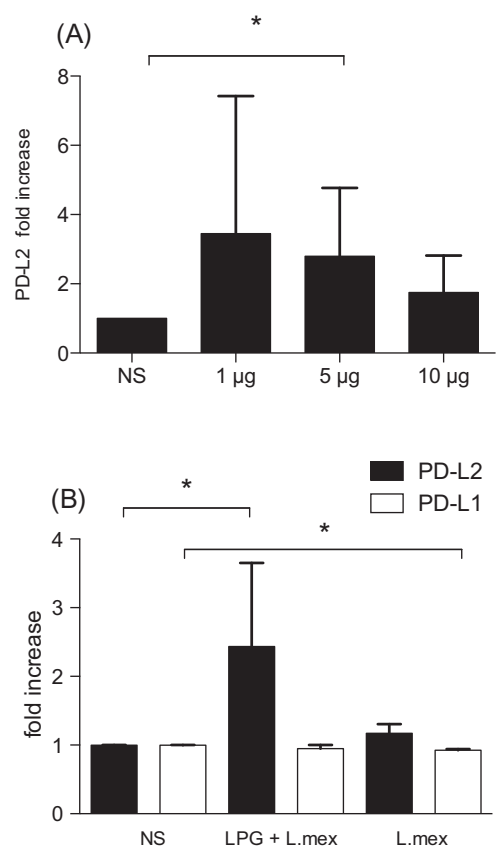

**Fig. 2.** LPG modulates the PD-L2 expression in peritoneal macrophages from healthy BALB/c. (A) Macrophages from peritoneal cavity were isolated and stimulated *in vitro* with different concentrations of LPG (1, 5 or 10 μg) during 24 h and stained with anti-PD-L2 mAb ($n = 3$). (B) Peritoneal macrophages were isolated and infected *in vitro* with *Leishmania mexicana* promastigotes at a 1:10 ratio (cells:parasites) or with parasites combined with 5 μg LPG. The expression of PD-L1 (white bars) and PD-L2 (black bars) was analyzed by flow cytometry. The bars represent normalized data of three separate experiments. Mean ± SD is shown. $^*p \le 0.05$ was considered significant.

### 3.3. L. mexicana LPG induces PD-1 expression in CD8$^+$ T cells of vaccinated mice

After finding that LPG exacerbated disease progression and modulated the PD-L2 expression in macrophages, we were interested in analyzing the effect exerted by LPG on spleen CD8$^+$ and CD4$^+$ T lymphocytes of mice immunized with two different doses of LPG. Vaccination with 10 or 100 μg LPG increased PD-1 expression in CD8$^+$ T cells. Re-stimulation of these cells *in vitro* with 1, 5 or 10 μg LPG maintained their elevated expression of PD-1 (Fig. 3A).

LPG had an opposite effect on CD137 expression in CD8$^+$ T cells. Mice vaccinated with 10 μg down-regulated their CD 137 expression by 20%, whereas vaccination with 100 μg decreased CD137 expression by 25% (Fig. 3B). Re-stimulation with 5 or 10 μg LPG further reduced CD137 in mice vaccinated with 10 μg, as compared to non-vaccinated controls (Fig. 3B).

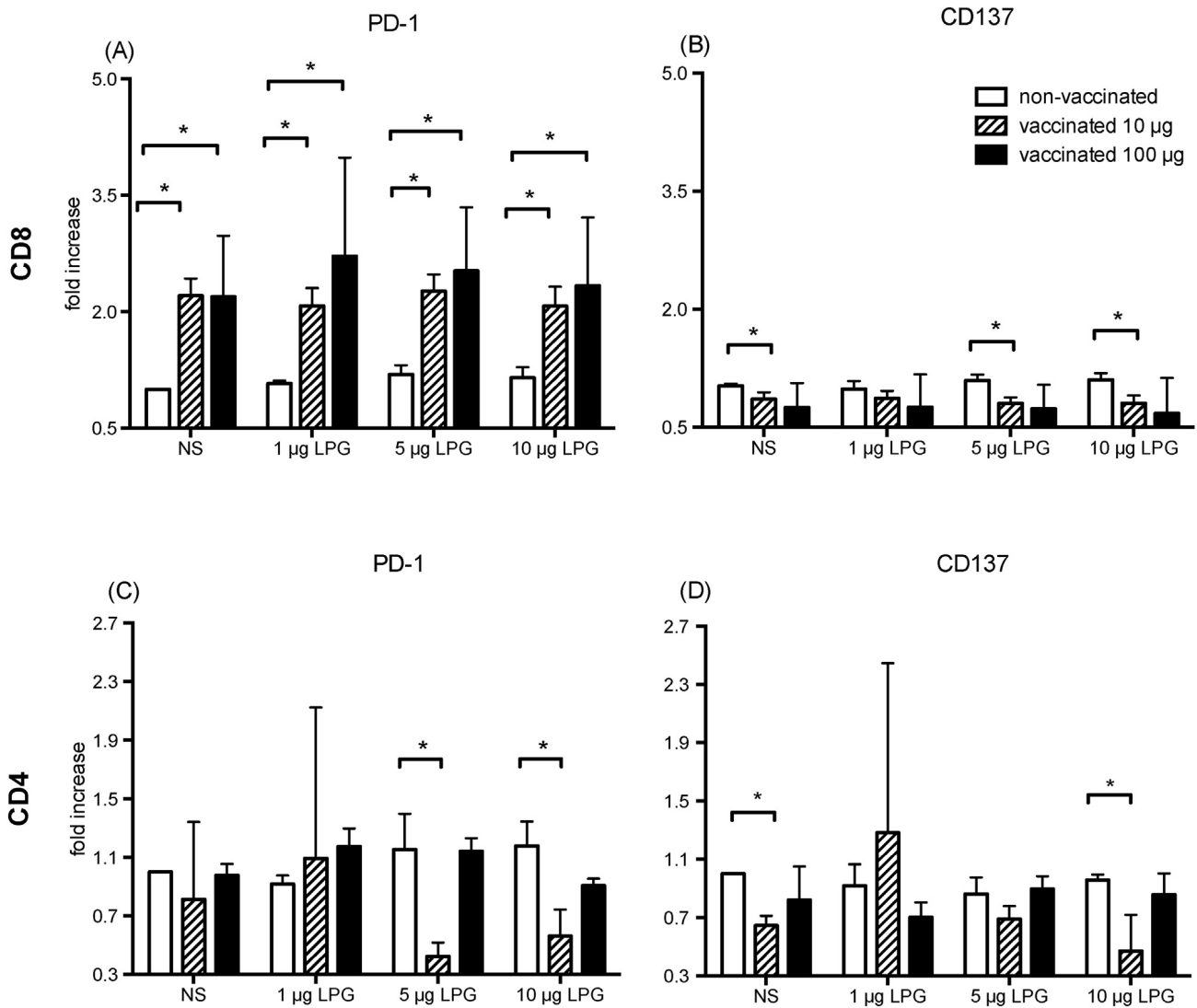

**Fig. 3.** PD-1 expression is dose-dependent in BALB/c vaccinated with *L. mexicana* LPG. Mice were vaccinated three times at 2-week intervals with 10 μg (stripped bars) or 100 μg LPG (black bars). Non-vaccinated mice were used as controls (white bars). Spleen T lymphocytes were obtained and re-stimulated *in vitro* with 1, 5 or 10 μg LPG during 24 h, fixed with paraformaldehyde and stained with anti CD3, CD4, CD8, PD-1 and CD137 antibodies. (A) PD-1 expression in CD8 T cells, (B) CD137 expression in CD8 T cells. (C) PD-1 expression in CD4 T cells, and (D) CD137 expression in CD4 T cells. The bars represent normalized data of three separate experiments. Mean ± SD is shown. * $p \leq 0.05$ was considered significant.

The analysis of CD4$^+$ T cells of mice vaccinated with 10 or 100 μg LPG showed no modification in the PD-1 expression. Yet *in vitro* re-stimulation with 5 or 10 μg LPG reduced PD-1 expression in CD4$^+$ cells of mice vaccinated with 10 μg, as compared to non-vaccinated controls (Fig. 3C). When analyzing the expression of CD137 in CD4$^+$ T cells, mice vaccinated with 10 μg mice showed a reduced expression, which diminished even more after these cells were re-stimulated *in vitro* with 10 μg LPG (Fig. 3D).

Together these data show that *L. mexicana* LPG negatively regulates CD8$^+$ cell activation by enhancing PD-1 expression and concomitantly reducing CD137 expressions, where the degree of the modulation depends upon the dose of LPG used for immunization as well as the dose of the subsequent stimulus. In contrast to CD8$^+$ T cells, vaccination with LPG had no inhibitory effect on CD4$^+$ T cells, since it did not modify their PD-1 expression and re-stimulation with LPG reduced their PD-1 expression. Thus, LPG vaccination seems to exert the inhibitory effect only on CD8$^+$ T cells, in a dose dependent fashion.

### 3.4. The expression of PD-1 in CD8$^+$ T lymphocytes of mice infected with L. mexicana is related to parasite load

To analyze whether parasite infection modulates PD-1 expression in T lymphocytes, BALB/c mice were infected in the earlobe dermis with $1 \times 10^4$ or $1 \times 10^5$ *L. mexicana* promastigotes. Mice were sacrificed prior to ulceration of the lesions. Splenocytes were isolated and re-stimulated *in vitro* with 1, 5 or 10 μg LPG during 24 h and PD-1 as well as CD137 were analyzed. We found that PD-1 expression is enhanced in CD8$^+$ T cells of mice infected with $1 \times 10^4$ (0.5-fold) or $1 \times 10^5$ (3.6-fold) parasites, as compared to CD8$^+$ T cells from non-infected mice (Fig. 4A). *In vitro* stimulation with all three doses of LPG showed the same high expression of PD-1.

The analysis of CD137 in CD8 T cells showed a 40% down-regulation in mice infected with $1 \times 10^4$ promastigotes, whereas mice infected with $1 \times 10^5$ promastigotes showed a similar expression as non-infected mice. *In vitro* re-stimulation with LPG did not alter CD137 expression (Fig. 4B).

CD4$^+$ lymphocytes showed a minimal increase in PD-1 expression after infections with either number *L. mexicana* parasites, and showed no changes despite secondary stimuli with LPG (Fig. 4C). Furthermore, the expression of CD137 in CD4$^+$ T cells of infected mice also remained unaltered. The only up-regulation of this activation marker was observed in CD4$^+$ T cells of mice infected with $1 \times 10^5$ parasites after they were re-stimulated *in vitro* with 5 μg LPG (Fig. 4D).

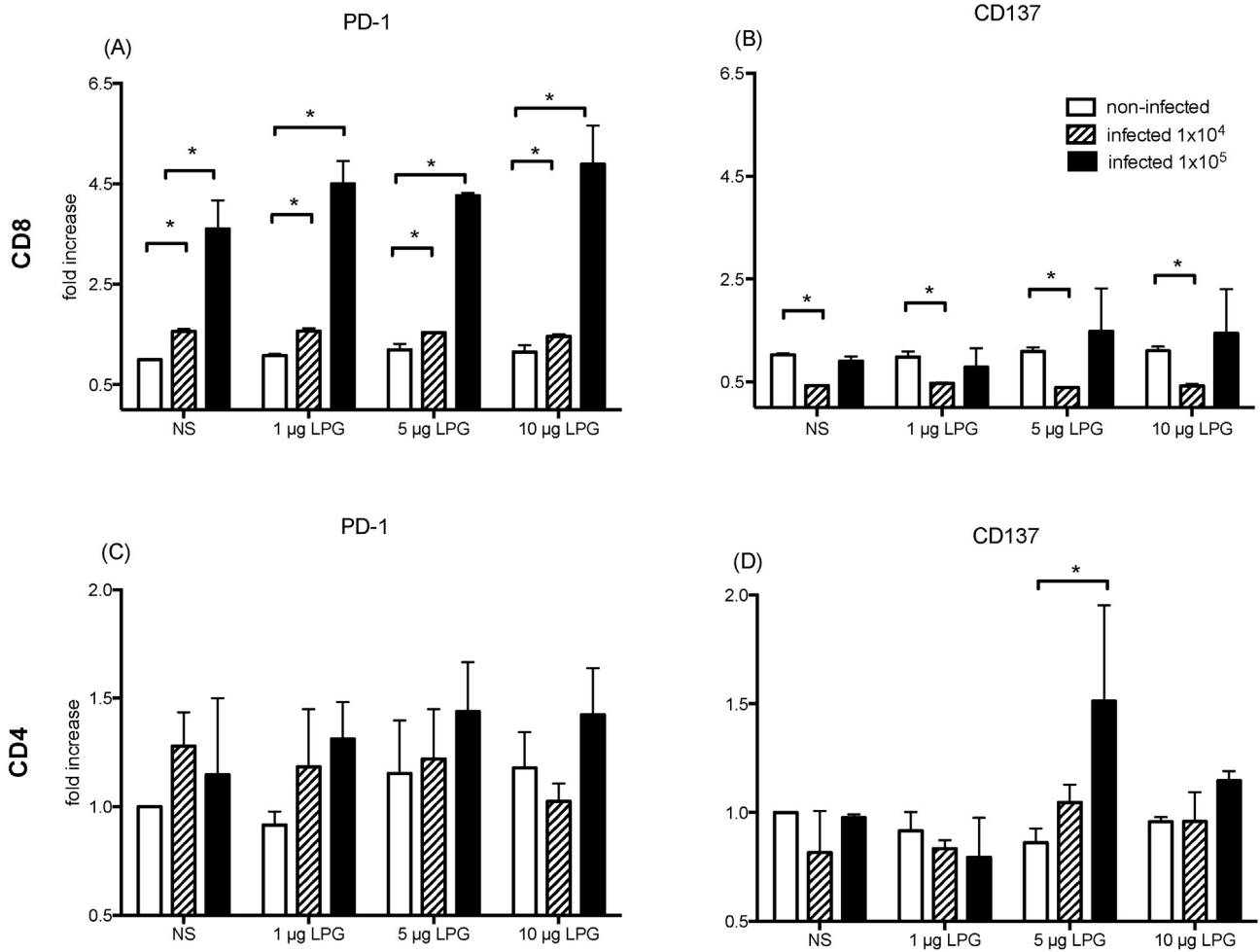

**Fig. 4.** *L. mexicana* infection promotes the PD-1 expression in CD8 T cells. Mice were infected in the ear dermis ear $1 \times 10^4$ (stripped bars) and a second group with $1 \times 10^5$ (black bars) *L. mexicana* promastigotes. Healthy BALB/c mice were used as a control group (white bars). Mice were euthanized before the lesions ulcerated. T cells from spleen were re-stimulated with 1, 5 or 10 μg LPG during 24 h, fixed with paraformaldehyde and stained with anti CD3, CD4, CD8, PD-1 and CD137 antibodies. (A) PD-1 expression in CD8 T cells, (B) CD137 expression in CD8 T cells, (C) PD-1 expression in CD4 T cells and (D) CD137 expression in CD4 T cells. The bars represent normalized data of three separate experiments. Mean ± SD is shown. * $p \leq 0.05$ was considered significant.

In conclusion these results show that *L. mexicana* infection induces significantly enhanced PD-1 expression only in CD8$^+$ T cells, in a dose-dependent fashion. The reduced expression of CD137 in association with the increased levels of PD-1 in these CD8$^+$ T cells seems to indicate that they resemble an exhausted phenotype. PD-1 is minimally expressed in CD4$^+$ cells during *L. mexicana* infections and not altered by *in vitro* LPG stimuli, showing that *L. mexicana* exerts a stronger inhibitory effect on CD8$^+$ T cells, as compared to CD4$^+$ T cells.

### 3.5. Vaccination with LPG or L. mexicana infection of mice regulates PD-L2 expression in spleen macrophages

Since vaccination with LPG immunomodulated CD8$^+$ T lymphocytes toward inhibition, we analyzed if immunization with different LPG concentrations or infection with different parasite numbers also modulated the expression of PD-L2 on spleen macrophages. Macrophages from mice vaccinated with 10 μg LPG and re-stimulated *in vitro* with 1 μg LPG, showed diminished expression of PD-L2 whereas vaccination with 100 μg LPG tended to increase the expression of PD-L2 in macrophages after receiving secondary stimuli with LPG (Fig. 5A).

Mice infected with $1 \times 10^4$ or $1 \times 10^5$ parasites down-regulated PD-L2 expression by 50% (Fig. 5B). Re-stimulation of macrophages from mice infected with $1 \times 10^4$ parasites with LPG always showed diminished expressions of this inhibitory marker, whereas those from mice infected with $1 \times 10^5$ parasites slightly increase their PD-L2 expression, albeit never reaching the levels expressed in cells of non-infected mice (Fig. 5B).

Together, these data show that *Leishmania* infections reduce PD-L2 expression in spleen macrophages and that this down-regulation persists despite secondary *in vitro* stimulation with LPG.

## 4. Discussion

Our data shed new light on the cause of enhanced disease progression after immunization with *Leishmania* LPG that has also been reported in the literature [16]. In an attempt to understand the underlying cause of this unsuccessful vaccination with LPG, we immunized mice with different concentrations of LPG and thereafter stimulated their spleen cells with various doses of LPG *in vitro* in an attempt to simulate a secondary exposure to LPG antigen, as T cells in controlling the parasite infection. The response of $CD4^+$ T cells was less clear.

PD-1 (programmed-death 1) receptor is related to CD28 and CTLA-4. It is inducible after T cell activation and down-regulates activated T cells [11]. Its ligands, PD-L1 and PD-L2, are up-regulated in APCs following activation [8]. PD-1 and PD-L2 may have distinctive roles in regulating Th-1 and Th-2 responses and reducing T cell proliferation by arresting the cell cycle [17,18]. This inhibitory receptor and its ligands have been studied in tumors, showing that the engagement of PD-1 with PD-L1 and PD-L2 attenuate T cell responses and help tumor cells escape immunosurveillance [19]. In chronic viral infections, suppressed $CD8^+$ T cell responses have been attributed to PD-1:PD-L1 interactions [20].

To the best of our knowledge, we here describe for the first time that suppressor receptor PD-1 is induced after vaccination with elevated doses of *Leishmania* LPG or with the infection with elevated amounts of *L. mexicana* promastigotes. This expression is specifically dominant on $CD8^+$ T lymphocytes possibly leading to a suppression of these cells that are critical in the control of leishmaniasis, both through IFN-$\gamma$ production, as well as in their cytotoxic effect against autologous *Leishmania*-infected macrophages [5,6]. These results call for a careful pre-immunization evaluation of potential vaccination candidates against *Leishmania*, since the induction of a suppressive effect can lead to detrimental blockage of the immune response, favoring a more virulent disease progression. These data open a new field of research in vaccine developments and provide a novel strategy for therapeutic intervention in leishmaniasis, where the blockade of PD-1 could represent a valuable approach for anti-*Leishmania* immunotherapy.

Our data also yield information on novel parasite evasion strategies, achieving $CD8^+$ T cell suppression, thereby eliminating one of the more powerful defense mechanisms against *L. mexicana* [13].

We conclude that vaccination models should assess whether PD-1 and/or PD-L2 are induced, that, far from activating $CD8^+$ T cells, it could lead to their inhibition. Additionally, during experimental models of *L. mexicana* infections, the parasite load must be taken into account, since it can have opposing effects on PD-1 expression in lymphocytes. This study provides insight into the regulatory pathways elicited in vaccine models using different antigen concentrations or during *Leishmania* infections with different parasite loads, showing that the outcome can be polarly opposed, leading to contradictory results.

### Acknowledgments

Maria Berenice Martínez Salazar was supported by a PhD fellowship from CONACyT and is a doctoral student of Programa de Doctorado en Ciencias Biomédicas, Universidad Nacional Autónoma de México (UNAM). The Project was financed by CONACyT—102155 and PAPIIT IN215212

*Conflict of interest*: The authors state that there is no conflict of interest.

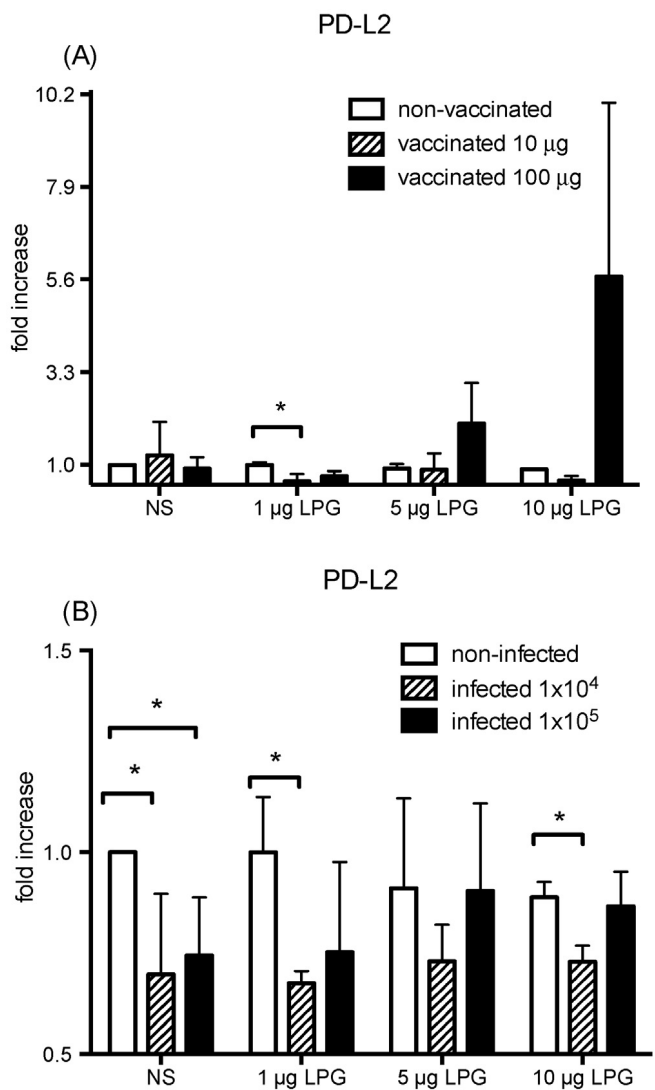

**Fig. 5.** Analysis of the PD-L2 expression in spleen macrophages from vaccinated (A) and infected mice (B). After vaccination with 10 or 100 μg LPG or infection with $1 \times 10^4$ or $1 \times 10^5$ *L. mexicana* promastigotes, macrophages were isolated from the spleen and re-stimulated with 1, 5 or 10 μg LPG during 24 h, fixed with paraformaldehyde and stained for F4/80 and PD-L2. The expression was analyzed by flow cytometry. (A) PD-L2 expression in vaccinated animals: white bars represent healthy mice, stripped bars represent mice vaccinated with 10 μg and black bars are mice vaccinated with 100 μg LPG. (B) PD-L2 expression in macrophages of mice infected with different parasite numbers. White bars represent healthy mice, stripped bars are mice infected with $1 \times 10^4$ and black bars are mice infected with $1 \times 10^5$ *L. mexicana* promastigotes. The bars represent normalized data of three separate experiments. Mean ± SD is shown. * $p \leq 0.05$ was considered significant.

would occur during a natural infection. Additionally, we infected mice with different *L. mexicana* numbers and also re-exposed their lymphocytes to a secondary challenge with LPG. We here show that immunization of BALB/c mice with LPG or infections with *L. mexicana* promastigotes enhances the expression of the inhibitory receptor PD-1 in $CD8^+$, whereas $CD4^+$ T cells remain unaltered. The increase of these inhibitory molecules in $CD8^+$ T cells acts in concert with their reduction of the activating molecule CD137, when these cells are confronted with a new challenge of LPG. These changes vary according to the amount of the LPG used for the vaccination and the parasite load during infection and they also vary according to the amount of parasite antigen (LPG) encountered by these cells after renewed exposure. The combination of these events possibly leads to a severe down-regulation of the functional capacity of $CD8^+$

### References

[1] Becker I, Salaiza N, Aguirre M, et al. *Leishmania* lipophosphoglycan (LPG) activates NK cells through toll-like receptor-2. Mol Biochem Parasitol 2003;130(2):65–74.
[2] Villasenor-Cardoso MI, Salaiza N, Delgado J, Gutierrez-Kobeh L, Perez-Torres A, Becker I. Mast cells are activated by *Leishmania mexicana* LPG and regulate the disease outcome depending on the genetic background of the host. Parasite Immunol 2008;30(8):425–34.
[3] Delgado-Dominguez J, Gonzalez-Aguilar H, guirre-Garcia M, et al. *Leishmania mexicana* lipophosphoglycan differentially regulates PKCalpha-induced oxidative burst in macrophages of BALB/c and C57BL/6 mice. Parasite Immunol 2010;32(6):440–9.
[4] Sacks D, Noben-Trauth N. The immunology of susceptibility and resistance to *Leishmania major* in mice. Nat Rev Immunol 2002;2(11):845–58.

[5] Belkaid Y, von Stebut E, Mendez S, et al. CD8+ T cells are required for primary immunity in C57BL/6 mice following low-dose, intradermal challenge with *Leishmania major*. J Immunol 2002;168(8):3992–4000.
[6] Uzonna JE, Joyce KL, Scott P. Low dose *Leishmania major* promotes a transient T helper cell type 2 response that is down-regulated by interferon gamma-producing CD8+ T cells. J Exp Med 2004;199(11):1559–66.
[7] Rhee EG, Mendez S, Shah JA, et al. Vaccination with heat-killed *Leishmania* antigen or recombinant leishmanial protein and CpG oligodeoxynucleotides induces long-term memory CD4+ and CD8+ T cell responses and protection against *Leishmania major* infection. J Exp Med 2002;195(12):1565–73.
[8] Latchman Y, Wood CR, Chernova T, et al. PD-L2 is a second ligand for PD-1 and inhibits T cell activation. Nat Immunol 2001;2(3):261–8.
[9] Loke P, Allison JP. PD-L1 and PD-L2 are differentially regulated by Th1 and Th2 cells. PNAS 2003;100(9):5336–41.
[10] Shin T, Yoshimura K, Shin T, et al. In vivo costimulatory role of B7-DC in tuning T helper cell 1 and cytotoxic T lymphocyte responses. J Exp Med 2005;201(10):1531–41.
[11] Carter L, Fouser LA, Jussif J, et al. PD-1:PD-L inhibitory pathway affects both CD4(+) and CD8(+) T cells and is overcome by IL-2. Eur J Immunol 2002;32(3):634–43.
[12] Salaiza-Suazo N, Volkow P, Tamayo R, et al. Treatment of two patients with diffuse cutaneous leishmaniasis caused by *Leishmania mexicana* modifies the immunohistological profile but not the disease outcome. Trop Med Int Health 1999;4(12):801–11.
[13] Hernandez-Ruiz J, Salaiza-Suazo N, Carrada G, et al. CD8 cells of patients with diffuse cutaneous leishmaniasis display functional exhaustion: the latter is reversed, in vitro, by TLR2 agonists. PLoS Negl Trop Dis 2010;4(11):e871.
[14] Joshi T, Rodriguez S, Perovic V, Cockburn IA, Stager S. B7-H1 blockade increases survival of dysfunctional CD8(+) T cells and confers protection against *Leishmania donovani* infections. PLoS Pathog 2009;5(5):e1000431.
[15] Pinheiro RO, Pinto EF, Lopes JR, Guedes HL, Fentanes RF, Rossi-Bergmann B. TGF-beta-associated enhanced susceptibility to leishmaniasis following intramuscular vaccination of mice with *Leishmania amazonensis* antigens. Microbes Infect 2005;7(13):1317–23.
[16] Pinheiro RO, Pinto EF, de Matos Guedes HL, et al. Protection against cutaneous leishmaniasis by intranasal vaccination with lipophosphoglycan. Vaccine 2007;25(14):2716–22.
[17] Greenwald RJ, Latchman YE, Sharpe AH. Negative co-receptors on lymphocytes. Curr Opin Immunol 2002;14(3):391–6.
[18] Carreno BM, Collins M. The B7 family of ligands and its receptors: new pathways for costimulation and inhibition of immune responses. Annu Rev Immunol 2002;20:29–53.
[19] Okazaki T, Honjo T. PD-1 and PD-1 ligands: from discovery to clinical application. Int Immunol 2007;19(7):813–24.
[20] Ascierto PA, Simeone E, Sznol M, Fu YX, Melero I. Clinical experiences with anti-CD137 and anti-PD1 therapeutic antibodies. Semin Oncol 2010;37(5):508–16.